\documentclass[journal=jacsat,manuscript=article]{achemso}
\usepackage[version=3]{mhchem} 
\usepackage{siunitx}
\usepackage[dvipsnames]{xcolor}
\usepackage{amssymb}
\usepackage{pifont}
\usepackage{wrapfig}
\usepackage{hyperref}
\usepackage{caption}
\usepackage{multirow}

\usepackage{amsmath} 

\usepackage{tabularx,colortbl}
\usepackage{capt-of}
\usepackage{adjustbox}

\author{Parisa Mollaei}
\affiliation[meche]
{Department of Mechanical Engineering, Carnegie Mellon University, 15213, USA}

\author{Danush Sadasivam}
\affiliation
{Department of Chemical Engineering, Carnegie Mellon University, 15213, USA}

\author{Chakradhar Guntuboina}
\affiliation[ece]
{Department of Electrical and Computer Engineering, Carnegie Mellon University, 15213, USA}

\author{Amir Barati Farimani}
\email{barati@cmu.edu}
\affiliation[meche]
{Department of Mechanical Engineering, Carnegie Mellon University, 15213, USA}
\alsoaffiliation[biomed]
{Department of Biomedical Engineering, Carnegie Mellon University, 15213, USA}
\alsoaffiliation[mld]
{Machine Learning Department, Carnegie Mellon University, 15213, USA}

\title[An \textsf{achemso} demo]
{IDP-Bert: Predicting Properties of Intrinsically Disordered Proteins (IDP) Using Large Language Models}

\abbreviations{IR,NMR,UV}
\keywords{American Chemical Society, \LaTeX}

\begin{document}
\begin{abstract}

\end{abstract}

Intrinsically Disordered Proteins (IDPs) constitute a large and structure-less class of proteins with significant functions. The existence of IDPs challenges the conventional notion that the biological functions of proteins rely on their three-dimensional structures. Despite lacking well-defined spatial arrangements, they exhibit diverse biological functions, influencing cellular processes and shedding light on the disease mechanisms. However, it is expensive to run experiments or simulations to characterize this class of proteins. Consequently, we designed an ML model that relies solely on amino acid sequences. In this study, we introduce IDP-Bert model, a deep-learning architecture leveraging Transformers and Protein Language Models (PLMs) to map sequences directly to IDPs properties. Our experiments demonstrate accurate predictions of IDPs properties, including Radius of Gyration, end-to-end Decorrelation Time, and Heat Capacity. 

\section{Introduction}

Within the complex frame of molecular biology, the long-standing belief that proteins' function strongly depends on their three-dimensional structures faces a compelling challenge due to the advent of a distinct class of biomolecules; Intrinsically Disordered Proteins (IDPs). IDPs challenge conventional structural norms, existing as dynamic ensembles of conformations that lack a well-defined spatial arrangement\cite{wright2015intrinsically, uversky2013unusual, oldfield2014intrinsically, dunker2001intrinsically, tompa2012intrinsically}. Despite their structural diversity, IDPs are often evolutionarily conserved across different species\cite{ahrens2017evolution, chen2006conservation}. This suggests that such proteins may play fundamental roles in biological processes and underscore their significance in the context of evolution\cite{handa2023perspectives, chen2006conservation2}. IDPs with their ability to adopt diverse shapes, are vital for essential cellular activities like gene expression and cell signaling\cite{bondos2022intrinsically, uversky2016dancing, chakrabarti2022intrinsically}, which is crucial for cell growth and response to stimuli. They often engage in complex interactions with other biomolecules\cite{uversky2019intrinsically, arai2024dynamics} that facilitate the formation of macromolecular complexes and cellular structures. These interactions are disrupted without IDPs, impacting cellular function. Thus, investigating IDPs properties can yield insights into disease mechanisms and avenues for therapeutic intervention\cite{saurabh2023fuzzy, rezaei2012intrinsically, wang2011novel, uversky2012intrinsically, ambadipudi2016targeting, joshi2015druggability}. However, performing experiments to characterize the properties of IDPs is expensive and is not feasible for large-scale screening. Moreover, running molecular dynamics (MD) simulations for analyzing IDPs' properties requires time-consuming and expensive computations. Therefore, a surrogate model is needed for rapid IDPs' properties prediction. 

In the last decade, machine learning (ML) models were successful in chemistry and building fast surrogate models for property prediction\cite{li2022graph, wang2022molecular, guntuboina2023peptidebert, yadav2022prediction, mollaei2023activity, kim2024gpcr, mollaei2023unveiling}. For proteins with well-known structures, several geometrical deep learning methods exist, such as graph neural networks (GNNs), the Equivariant Graph Neural Nets (EGNN), etc\cite{liao2023equiformerv2, wang2023graph, wang2022molecular, li2022graph}. Inspired by these advances, we developed an ML model to predict the IDPs properties. Given the absence of structural features in IDPs, our model relies primarily on the arrangement of amino acids in protein sequences. Although ML models such as Bag of Words and fingerprinted features have been used to model the IDPs properties\cite{goldberg2022neural}, they often fall short in prediction accuracy. These methods lack the ability to adequately capture the sequential relationships between amino acids, particularly the long-range connections, which are crucial for accurately predicting the proteins' properties. In previous works, Lotthammer et al\cite{lotthammer2024direct}. developed ALBATROSS by simulating synthetic and natural IDP sequences using the Mpipi-GG force field and training bidirectional recurrent neural networks (BRNN-LSTM)\cite{schuster1997bidirectional} for sequence-to-ensemble property prediction. Tesei et al\cite{tesei2024conformational}. trained SVR models to predict structural properties including interaction energy maps, asphericity, prolateness, and an estimate of the conformational entropy per residue using sequence descriptors, with optimal hyperparameters determined via grid searches. 

The advent of Transformers\cite{vaswani2017attention, wolf2020transformers} and Protein Language Models (PLMs)\cite{thirunavukarasu2023large} has catalyzed the development of deep learning architectures in modeling amino acid sequences as analogous to words and sentences in languages\cite{defresne2021protein, guntuboina2023peptidebert, kim2024gpcr}. The attention mechanism inherent in PLMs enables them to adeptly capture both immediate and intricate connections among various elements of textual data\cite{vaswani2017attention}.  Specifically, combination of pre-training with transformers leads to creation of "BERT"\cite{devlin2019bert} which is a powerful language encoder. These advancements have sparked a renaissance in bioinformatics, as protein sequences, much like languages, reveal intricate interactions among their constituent amino acids. With the capabilities offered by PLMs and Transformers, we are able to explore the contributions of amino acids to the features of proteins\cite{guntuboina2023peptidebert, kim2024gpcr}. Through the utilization of pre-trained models like ProtBERT\cite{elnaggar2021prottrans}, we fine-tuned it to accurately forecast the IDPs' properties. The ProtBERT model has been trained on extensive datasets of protein sequences, enabling them to learn comprehensive representations of protein sequences\cite{elnaggar2021prottrans}. Previous studies have demonstrated the efficacy of learning algorithms, particularly PLMs, in deciphering the language of proteins and uncovering their intrinsic characteristics\cite{guntuboina2023peptidebert, ock2023catalyst, ramadoss2023pre, kim2024gpcr, patil2023forecasting}. Our experiments revealed that the PLMs can also accurately predict structural, dynamic, and thermodynamic properties of IDPs, including Radius of Gyration, end-to-end Decorrelation Time, and Heat Capacity.

The Radius of Gyration provides information about the compactness and spatial extent of the protein structure which is defined as the root mean square distance of all atoms in a protein from their common center of mass. For IDPs, which lack a well-defined three-dimensional structure, the Radius of Gyration provides insight into their conformational flexibility and structural heterogeneity\cite{lobanov2008radius}. Radius of Gyration helps infer the impact of factors like pH, temperature, or binding partners on the protein conformations. Insights into IDP conformational changes that are essential for interactions in signaling pathways and other functions can be useful in designing molecules that target or stabilize specific IDP conformations. The end-to-end Decorrelation Time measures the timescale over which the positional correlation between two ends of the protein chain diminishes, providing insight into the dynamics and flexibility of the protein\cite{trement2014conservative}. This property is indicative of the speed of conformational changes and hence can determine how quickly IDPs can form or break interactions with other molecules. This is crucial for their roles in signaling and regulation, where rapid response times are often necessary. The Heat Capacity reflects the protein's ability to absorb and dissipate heat, which is related to its stability and structural dynamics. IDPs' thermal stability and heat capacity are significantly different compared to structured proteins which highlights their ability to adopt multiple conformations and undergo conformational changes easily\cite{uversky2017paradoxes}. Variations in heat capacity can reveal changes in the conformational states and stability of IDPs under different conditions. Heat capacity is essential for understanding thermal transitions in IDPs, such as protein folding and are hence, key to comprehending the functional states and activity of IDPs. Gaining insight into these properties yields valuable information regarding IDPs' biological functionalities and versatility in molecular interactions.

\section{Methods}

\begin{figure}[t!]
     \centering
     \includegraphics[width=0.75\linewidth]{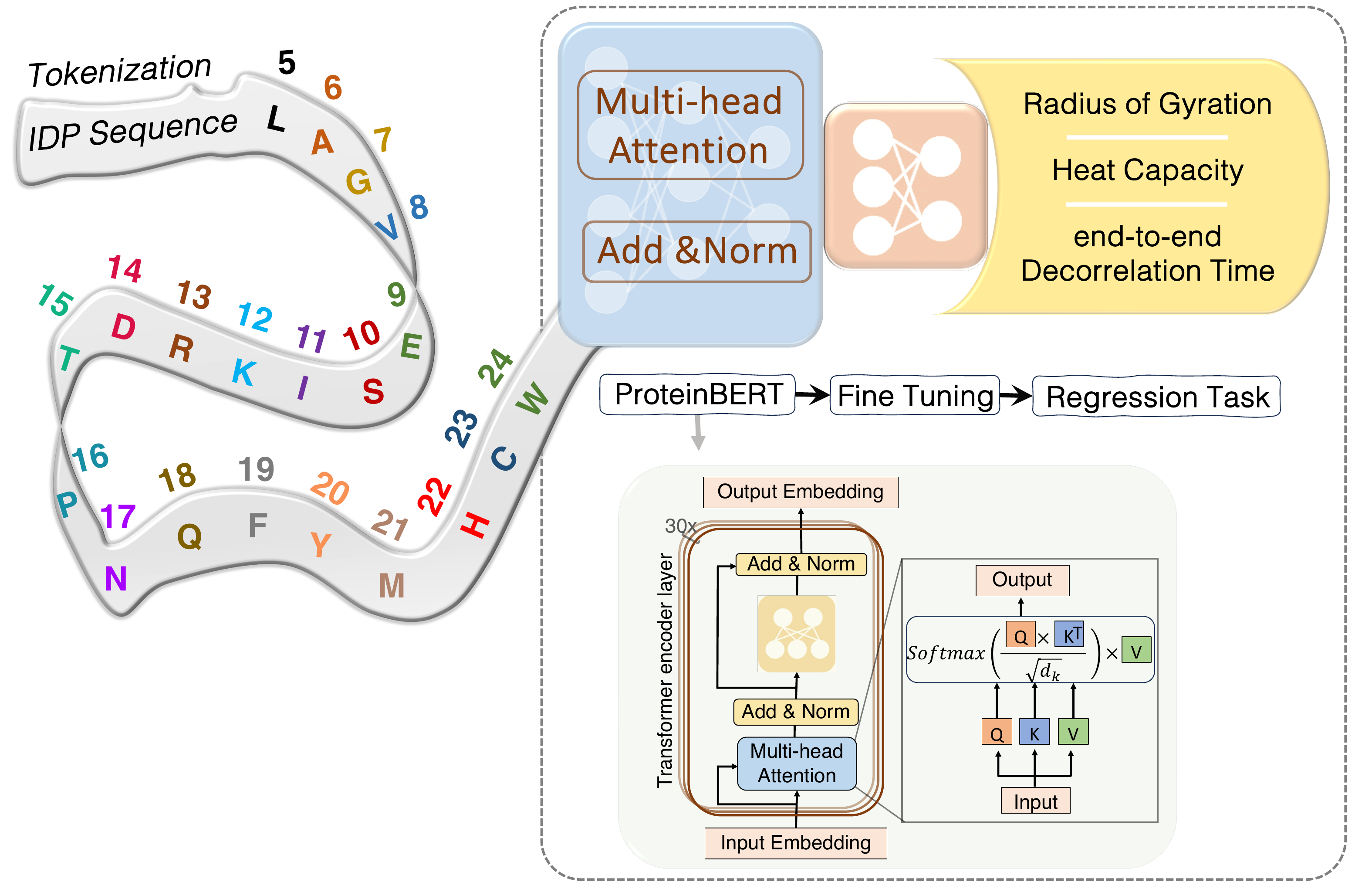}
     \caption{The overall IDP-Bert framework illustrates the preprocessing of input data, tokenization, integration with ProtBERT, fine-tuning process, and regression tasks aimed at predicting the Radius of Gyration, end-to-end Decorrelation Time, and Heat Capacity of IDPs.}
     \label{fig:idp_model}
 \end{figure}

The schematic representation of IDP-Bert's overall architecture is depicted in Figure\ref{fig:idp_model}. IDP-Bert utilizes ProtBERT\cite{elnaggar2021prottrans}, a transformer-based model with 16 attention heads and 30 hidden layers as its backbone. ProtBERT is pre-trained on Big Fantastic Database (BFD)\cite{jumper2021highly, steinegger2019protein, steinegger2018clustering}, a vast protein sequence corpus comprising over 217 million unique sequences. An adaptation of the original BERT\cite{devlin2018bert} developed by Google, ProtBERT adopts the encoder segment structure of the transformer model, featuring multiple sequential layers of an attention-feed-forward network, as illustrated in Figure\ref{fig:idp_model}. Within the attention mechanism, each token is encoded into an input embedding and then transformed into keys, queries, and values. Keys and queries are combined through matrix multiplication to construct the attention map, which subsequently undergoes a softmax function\cite{bridle1990probabilistic} to produce a probability distribution. Following this, the resulting distribution is employed to scale (multiply) the value vectors. The feed-forward layer within each Transformer layer aids in capturing intricate patterns within the input, while the attention mechanism encodes relationships among various tokens. The multi-head attention structure partitions the input across multiple parallel attention layers, or "heads", allowing each head to independently specialize in detecting diverse patterns and relationships\cite{vaswani2017attention}. This configuration of the Transformer encoder in ProtBERT enables the model to learn context-aware representations of amino acids in protein sequences. One significant advantage of ProtBERT is its superior performance across multiple benchmarks\cite{elnaggar2021prottrans}. For the fine-tuning process, a series of experiments were conducted to determine the optimal configuration for the ProtBERT backbone and the architecture for the regression head, resulting in the selection of a design comprising two fully connected layers (refer to Supplementary Information section Experiments).

\subsection{Data preprocessing and model}

For the dataset used in this study, the MD simulations were conducted by Patel et al.\cite{patel2022featurization}, employing the hydropathy scale (HPS) model via the LAMMPS simulation package.  Prior to the simulations, an energy minimization step was performed, followed by a $10^9$ fs run with time steps of 10 fs. The system was maintained at 300 K using the Langevin thermostat with a damping constant of 1 ps. Thermodynamic properties necessary for calculating the heat capacity ($C_v$) were acquired at 100 ps intervals, while atom coordinates, crucial for determining the radius of gyration ($R_g$) and the decorrelation time ($\tau_N$), were obtained at 5 ps intervals. The following equations were used to calculate $R_g$, $C_v$, and $\tau_N$:

\begin{equation}
R_g := \langle R^2 \rangle^{1/2} = \left( \frac{\sum_{i=1}^{N} \langle (R_i - \langle R_{CM} \rangle)^2 \rangle}{N} \right)^{1/2}
\end{equation}

where $N$ represents the total number of atoms, while $R_i$ and $R_{CM}$ denote the positions of atoms and the center of mass of all atoms within the system, respectively.

\begin{equation}
C_v := \langle C_v \rangle = \frac{\langle E^2 \rangle - \langle E \rangle^2}{k_B T^2}
\end{equation}

where $E$ is the total internal energy of the system, and $k_B$ is the Boltzmann constant.

\begin{equation}
\tau_N := \langle \tau_N \rangle = \int_{0}^{\infty} \langle \delta R(t) \delta R(0) \rangle dt, \quad \delta R(t) = R_{i=N}(t) - R_{i=1}(t)
\end{equation}

At a given time t, the $R_{i=N}(t)$ and $R_{i=1}(t)$ represent the end positions of the polymer. The integral was approximated by fitting the end-to-end time autocorrelation function to a Kohlrausch–Williams–Watts (KWW) function and then performing an analytical integration.

The dataset contains properties derived from simulations for 2585 IDPs, each with a degree of polymerization ranging from 20 to 600 amino acids\cite{patel2022featurization}. These IDPs are characterized as linear and stochastic polymers due to their sequences. The sequences were sourced from version 9.0 of the DisProt database\cite{hatos2020disprot, piovesan2017disprot} to ensure uniqueness. Properties of the IDPs were computed using molecular dynamics simulations at 300 K with the LAMMPS simulation package\cite{thompson2022lammps}, employing the improved hydropathy scale (HPS) CG model\cite{regy2021improved}. Key properties, including Radius of Gyration ($R_g$), end-to-end Decorrelation Time ($\tau_N$), and Heat Capacity ($C_v$), were determined for use as labels in regression tasks. The labels were subjected to a data transformation method known as Yeo-Johnson transformation\cite{yeo2000new} to reduce skewness and non-normality in their distribution, thereby improving the performance of the model. This technique introduces a power transformation, facilitating a closer approximation to a normal distribution.
This dataset comprising 2585 IDPs was randomly divided into three distinct subsets; train, validation, and test sets, each non-overlapping with the others. MMSeqs2\cite{Steinegger2017} was used to perform a sequence similarity check of the sequences in the dataset with a similarity threshold of 0.8. None of the sequences were clustered, verifying that the sequences in the dataset are not more than 80\% similar.

To transform raw protein sequences into a format that can be effectively processed by ProtBERT, they need to be tokenized (Figure\ref{fig:idp_model}). Each amino acid is represented by a number (known as a token) in the range of 5 to 24 (inclusive), with numbers 0 through 4 being reserved for special tokens (0: [PAD], 1: [UNK], 2: [CLS], 3: [SEP], and 4: [MASK]). These tokens were fed to IDP-Bert and three separate models were trained (one for each IDP property). We employed the Mean Squared Error (MSE) loss to train our model over 5 epochs and used $R^2$ scores to assess the performance of our model. We used a modified configuration for ProtBERT with 16 hidden layers (each with a size of 256), a hidden layer dropout probability of 0.15, and 16 attention heads. We attach a regression head on top of this which consists of 2 fully connected layers, each followed by a ReLU activation. The model was optimized using the AdamW optimizer\cite{loshchilov2017decoupled} with an initial learning rate of $1*10^{-5}$. ReduceLROnPlateau scheduler\cite{ReduceLROnPlateau} was used to improve the convergence of the loss. The scheduler parameters were set to reduce the learning rate by a factor of 0.1 if there was no improvement in the loss over 3 epochs.

Upon dividing the dataset into buckets based on the length of the sequences, we noticed that buckets with fewer data points exhibited higher average MSE values (Figure\ref{fig:hist}a-c). To address this and enhance our model's generalizability, we used a data augmentation\cite{van2001art} technique known as oversampling. Oversampling involves artificially augmenting the number of data points in the under-represented class by duplicating them. Specifically, we determined the bucket with the highest number of points denoted as N-max. For each bucket with a lower number of points, denoted as N-bucket, we calculated the multiplier by rounding the ratio of N-max/N-bucket  down to the nearest integer. Subsequently, each bucket with fewer points was duplicated to increase its size, resulting in a balanced dataset (Figure\ref{fig:hist}d-f) with a similar number of data points across all buckets. It is important to note that this augmentation is only applied to the training set.

\section{Results and discussion}

\begin{table}[t!]
\begin{adjustbox}{width=0.6\columnwidth,center}
\begin{tabular}{ |c|c|c| } 
\hline
IDP’s property & Average of ${R^2 \pm SD}$ \\
\hline
Radius of Gyration & ${0.9881 \pm 0.0018}$ \\
end-to-end Decorrelation Time & ${0.9713 \pm 0.0072}$ \\ 
Heat Capacity & ${0.9645 \pm 0.0042}$ \\ 
\hline
\end{tabular}
\caption{\label{tab:acc}Performance evaluation of IDP-Bert model. Average ${R^2}$ Values for Radius of Gyration, End-to-End Decorrelation Time, and Heat Capacity Predictions with Standard Deviations (SD).}
\end{adjustbox}
\end{table}

The average $R^2$ values, along with their corresponding standard deviations, were computed after training the model five times, each with a distinct train-test split. These metrics were utilized to evaluate the predictive performance of the IDP-Bert model in estimating property values. Table \ref{tab:acc} shows the average $R^2$ is 0.9881 for the Radius of Gyration, indicating approximately 98.81\% of the variance in Radius of Gyration can be explained by the independent variable(s) modeled by IDP-Bert. The standard deviation associated with this is 0.0018. The average $R^2$ for end-to-end Decorrelation Time and Heat Capacity are 0.9713 and 0.9645, respectively, with corresponding standard deviations of 0.0072 and 0.0042 (Table \ref{tab:acc}). These results suggest a strong relationship between the features used in IDP-Bert model and the IDP's properties, underscoring a high degree of predictability of the model.

\begin{table}[t!]
    \begin{adjustbox}{width=0.5\columnwidth,center}
    \begin{tabular}{ |c|>{\centering\arraybackslash}p{3.5cm}| } 
    \hline
    Property & Average inference time per sample (s) \\
    \hline
    Radius of Gyration & 0.0254 \\
    Heat Capacity & 0.0384 \\
    Decorrelation Time & 0.0265 \\
    \hline
    \end{tabular}
    \end{adjustbox}
    \caption{\label{tab:inference_time}The average time taken for inference for the trained IDP-Bert models to obtain predictions for a single protein sequence.}
\end{table}

The time taken for inference was noted during the testing stage of each model, and the average inference time taken per sample is reported in Table \ref{tab:inference_time}. The time taken to obtain the predictions for the entire testing dataset of 522 protein sequences was recorded, and an average was computed across all samples to determine the inference time per sample. This was done to obtain a mean inference time over the varying sequence lengths present in the dataset.


\begin{table}[t!]
\begin{adjustbox}{width=0.65\columnwidth,center}
\begin{tabular}{ |c| >{\centering\arraybackslash}p{3cm} | >{\centering\arraybackslash}p{3cm} | } 
\hline
Property & Baseline mean $R^2$ & IDP-Bert mean testing $R^2$ \\
\hline
Radius of Gyration & 0.9859 & 0.9881 \\
Heat Capacity & 0.8823 & 0.9645 \\
Decorrelation Time & 0.7828 & 0.9713 \\
\hline
\end{tabular}
\end{adjustbox}
\caption{\label{tab:baseline_comparison}Comparison of performance of IDP-Bert models with the best baseline models.}
\end{table}

The results of the IDP-Bert model were compared with the best pre-existing models for predicting these as discussed by Patel et al \cite{patel2022featurization} in Table \ref{tab:baseline_comparison}. In their study, the best performing model for Radius of Gyration utilized scaled force field fingerprints, achieving a mean $R^2$ of 0.9859. For Decorrelation Time, their best model also used scaled force field fingerprints, with a mean $R^2$ of 0.7828. For Heat Capacity, the best model used Morgan fingerprints, attaining a mean $R^2$ of 0.8823. Our IDP-BERT framework surpasses these benchmarks, yielding mean testing $R^2$ scores of 0.9881 for Radius of Gyration, 0.9713 for Decorrelation Time, and 0.9645 for Heat Capacity.

 \begin{figure}[t!]
     \centering
     \includegraphics[width=0.95\linewidth]{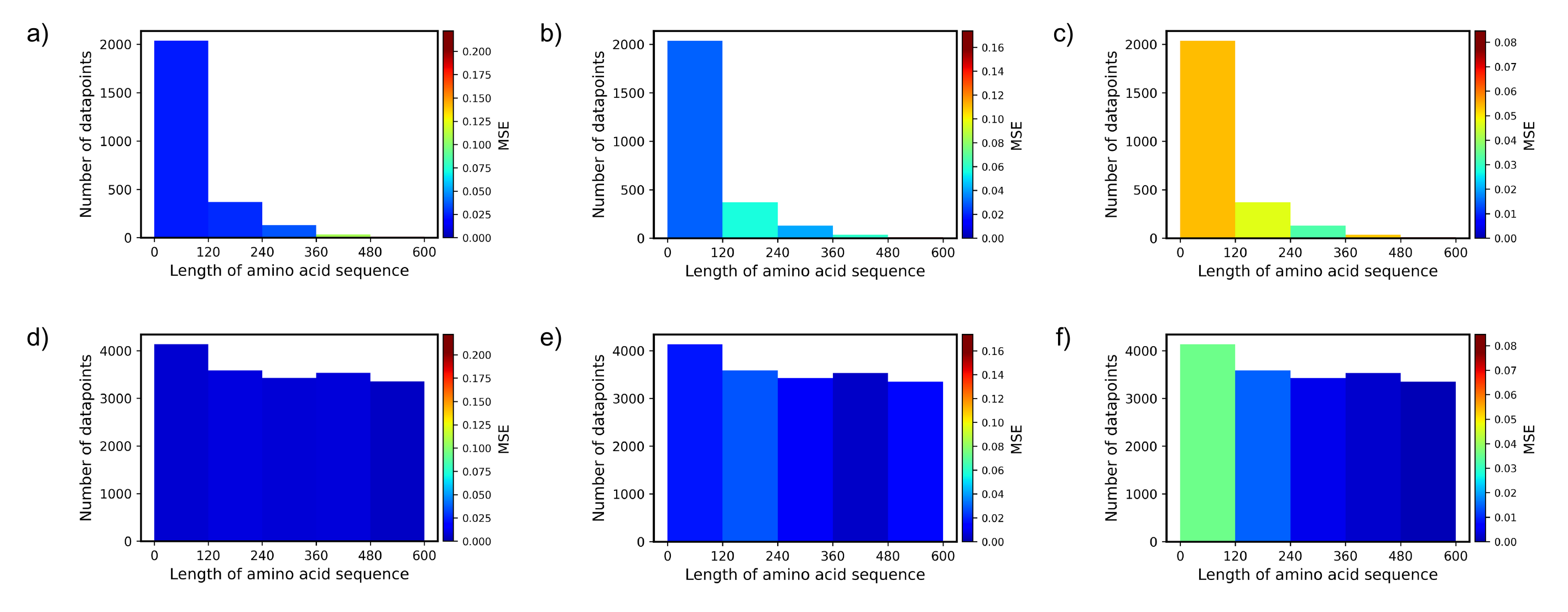}
     \caption{Distribution of original and augmented IDP samples based on the length of their amino acid sequences. The Mean Squared Error (MSE) for Radius of Gyration is depicted for both (a) original and (d) augmented datasets, followed by the MSE for end-to-end Decorrelation Time shown for (b) original and (e) augmented datasets. Lastly, the MSE for Heat Capacity is presented for (c) original and (f) augmented datasets.}
     \label{fig:hist}
 \end{figure}

To evaluate the effect of augmentation process on the results, we mapped the MSE values onto each range of sequence length for both the original and augmented dataset across each IDP property: Radius of Gyration (Figure\ref{fig:hist} (a),(d)), end-to-end Decorrelation Time (Figure\ref{fig:hist}(b),(e)), and Heat Capacity (Figure\ref{fig:hist} (c),(f)). This information provides insights into the impact of both number of samples and length of IDP sequences on the accuracy of the predicted values. As depicted in Figure\ref{fig:hist}a,b, it is evident that the MSE values for large sequences exhibit notably higher values. This may be attributed to several reasons, such as larger IDP sequences tend to have more complex structures and conformational variability that make predicting IDP properties more challenging. Furthermore, the limited dataset of longer sequences in the training dataset can lead to poorer performance of IDP-Bert model. However, data augmentation improved the model's performance for the large sequences (compare Figure\ref{fig:hist}(b) with (e) or Figure\ref{fig:hist}(c) with (f)).

 \begin{figure}[t!]
     \centering
     \includegraphics[width=\linewidth]{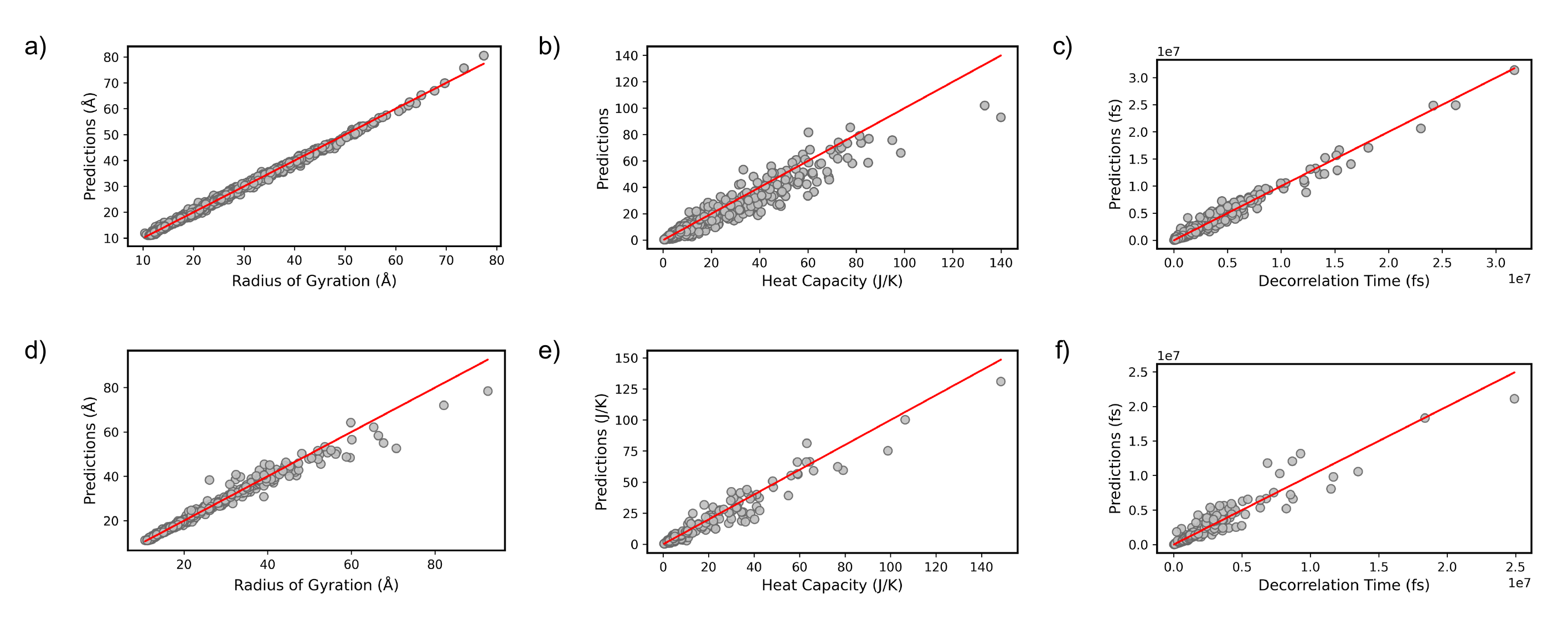}
     \caption{The correlation between the predicted and ground truth values, accompanied by their corresponding $R^2$ coefficients for the train and test sets from a single run. The Radius of Gyration (a) and (d), end-to-end Decorrelation Time (b) and (e), and Heat Capacity (c) and (f).}
     \label{fig:corr}
 \end{figure}
 
With an emphasis on reliability and efficiency of IDP-Bert model in capturing the essential features of IDPs, our analysis reveals a robust correlation between the predicted properties and corresponding ground truth values in both train and test sets from a single run (Figure\ref{fig:corr}). The slight differences between $R^2$ values in test and train sets indicate that the model has learned the underlying pattern within the data without memorizing the noise in the training set, allowing it to make accurate predictions on unseen data. Additionally, it validates the generalizability of our model across different datasets.

\begin{figure}[t!]
     \centering
     \includegraphics[width=\linewidth]{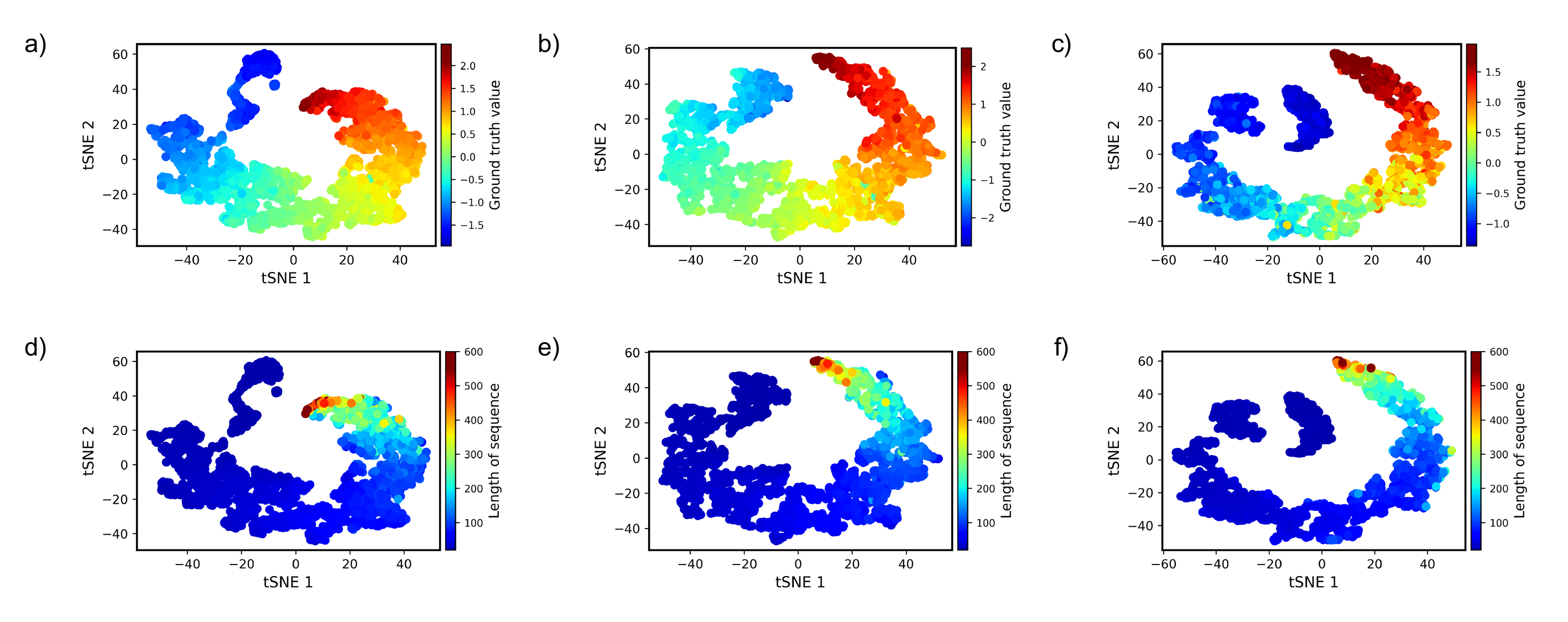}
     \caption{Ground truth values and sequence lengths mapped onto the t-SNE representation for Radius of Gyration (a,d), end-to-end Decorrelation Time (b,e), and Heat Capacity (c,f)}
     \label{fig:tsne}
 \end{figure}

Building upon the strong correlation observed between predicted and ground truth values, we further explored the features extracted from the penultimate layer of the IDP-Bert framework (within fine-tuning part in Figure \ref{fig:idp_model}). Utilizing these features, we generated t-distributed Stochastic Neighbor Embedding (t-SNE) visualization to gain deeper insights into the underlying relationships in the data. The t-SNE is a technique for reducing the dimensionality of data while preserving local structure, making it useful for visualizing complex datasets in lower dimensions\cite{van2008visualizing}. 
In our analysis, the t-SNE interpretation reflects the relationship between IDP properties, sequence lengths, and their structural and functional characteristics. Figure\ref{fig:tsne}(a), (b), and (c) show the ground truth values mapped onto the t-SNE representations for Radius of Gyration, end-to-end Decorrelation Time, and Heat Capacity, respectively. Notably, the figure exhibits a discernible pattern, gracefully transitioning from minimum to maximum values of the ground truth across the representations. This smooth progression suggests that the fine-tuned results used for the t-SNE algorithm have effectively preserved the underlying structure and relationships in the original high-dimensional data. Furthermore, the patterns observed in the t-SNE representations suggest that there may be inherent structural similarities among IDPs with similar properties. To justify this, we mapped the lengths of sequences onto the t-SNE representations (as shown in Figure \ref{fig:tsne} (d), (e), and (f)). The comparison of ground truth values and sequence lengths on the t-SNE representations (Figure \ref{fig:tsne}) verifies that longer sequences, which often exhibit greater structural complexities, are associated with higher property values than shorter ones. 

\begin{figure}[t!]
     \centering
     \includegraphics[width=0.95\linewidth]{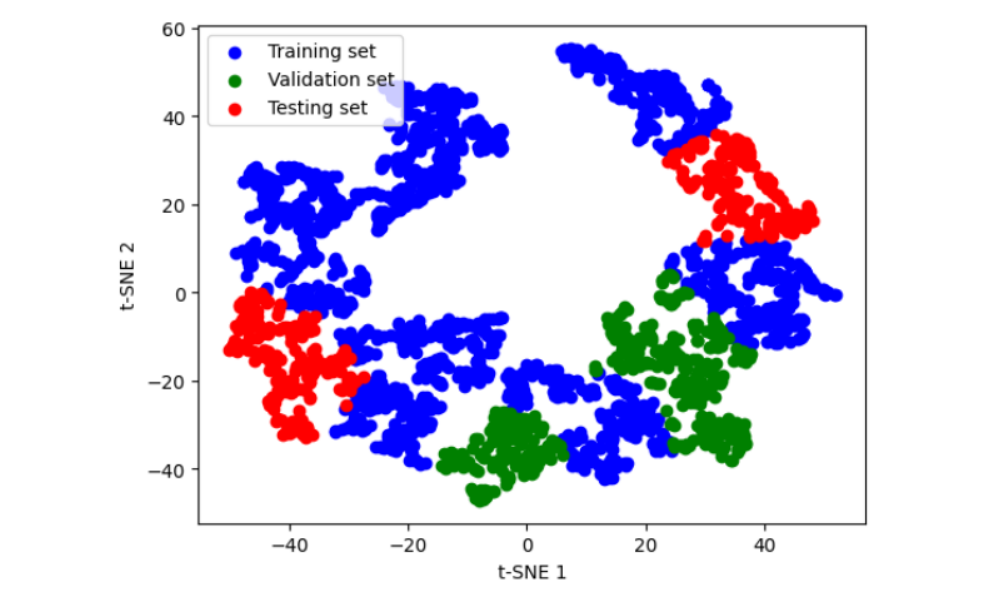}
     \caption{KMeans clustering of the t-SNE embeddings of the intermediate linear layer outputs from the IDP-Bert model. The 15 clusters were randomly assigned as Training, Validation and Testing sets.}
     \label{fig:tsne_clustering}
\end{figure}

To investigate the impact of clustering on model training, we formed clusters using KMeans clustering on the t-SNE embeddings of the intermediate linear layer outputs from the model. This resulted in 15 clusters, which were then randomly assigned to training, validation, or test sets. This clustering can be seen in Figure \ref{fig:tsne_clustering}. Specifically, 9 clusters were used for training, and 3 clusters each for validation and testing, maintaining an approximate 60:20:20 split. These results have been reported in Table \ref{tab:clustering_results}.

\begin{table}[t!]
\begin{adjustbox}{width=\columnwidth,center}
\begin{tabular}{ |p{5cm}|c|c|c|c| }
\hline
Clustering & Property & Training $R^2$ & Validation $R^2$ & Testing $R^2$ \\
\hline
\multirow{3}{*}{\parbox[t]{5cm}{The 15 clusters were randomly assigned as training, validation, or testing sets.}} & Radius of Gyration & 0.9655 & 0.7138 & 0.9472 \\ [3pt]
& Heat Capacity & 0.9502 & 0.6800 & 0.9561 \\ [3pt]
& Decorrelation Time & 0.9590 & 0.5866 & 0.9470 \\[3pt]
\hline
\multirow{4}{*}{\parbox[t]{5cm}{The training, validation, and testing sets were proportionally sampled from each cluster in a 60:20:20 ratio.}} & Radius of Gyration & 0.9549 & 0.9710 & 0.9616 \\[9pt]
& Heat Capacity & 0.9349 & 0.9528 & 0.9490 \\[9pt]
& Decorrelation Time & 0.9511 & 0.9634 & 0.9614 \\[9pt]
\hline
\end{tabular}
\end{adjustbox}
\caption{\label{tab:clustering_results}Training, Validation and Testing $R^2$ values for Radius of Gyration, Heat Capacity, and Decorrelation Time for two different data splits based on the clusters obtained from KMeans clustering.}
\end{table}

We observed that while the model performed well on the training and test sets, its performance on the validation set was not as strong. This can be attributed to the model's lack of exposure to that region of the dataset during training, affecting its ability to interpolate. Additionally, we compared these results with a scenario where sampling points from each cluster in a 60:20:20 ratio was used for training. The results from this study are also reported in Table \ref{tab:clustering_results}. In this case, the model saw a diverse set of points from all regions of the dataset during training and performed better across all three sets (training, validation, and test).


\section{Conclusion}
This study introduces IDP-Bert model, a fine-tuned protein language model (PLM), to effectively predict some of the Intrinsically Disordered Protein (IDPs) properties. Given the inherent lack of well-defined structures in IDPs, this model is designed to decipher the language of IDPs and precisely predict their intrinsic characteristics based solely on amino acid sequences. Through our experiments, accurate predictions of structural, dynamic, and thermodynamic properties, including Radius of Gyration, end-to-end Decorrelation Time, and Heat Capacity, were achieved. Additionally, analysis of attention weights assigned to individual amino acids within IDPs illuminates their contributions to determining protein properties that facilitate protein engineering and drug design efforts.

\section{Data and software availability}
The necessary information containing the codes and data for downstream tasks used in this study is available here: \url{https://github.com/DanushSadasivam/IDP-BERT}

\section{Supporting Information file}
Experimental details, including six experiments with varying hyperparameters such as hidden layer size, number of attention heads, and the number of hidden layers, and their corresponding R² values for train, validation, and test sets. Analysis of the contribution of sequence length in predicting properties, with modified training procedures and the resulting performance metrics. Protein taxonomy data obtained from a similarity search on the UniProt database, including the UniProt IDs of the similarity hits, similarity fractions, and taxonomy information in a CSV file named "taxonomy results.csv".

\begin{acknowledgement}

This work is supported by the Center for Machine Learning in Health (CMLH) at Carnegie Mellon University and a start-up fund from the Mechanical Engineering Department at CMU. 

\end{acknowledgement}

\bibliography{reference}


\end{document}


\maketitle

\section{Experiments}

Table \ref{tab:experiments} presents six experiments showcasing diverse hyperparameters, including the number of epochs, batch size, size and number of hidden layers, number of attention heads, number of layers in the head, and dropout percentage. Following these variations, $R^2$ results for the train, validation, and test sets are provided. The average $R^2$ for the test sets serves as the performance metric for the IDP-Bert model in predicting each of the Radius of Gyration, end-to-end Decorrelation Time, and Heat Capacity properties.


\begin{table}[t!]
\renewcommand{\arraystretch}{1.2}
\begin{adjustbox}{width=\columnwidth,center}
\begin{tabular}{|c|c|c|c|c|c|c|c|c|c|c|}
\hline
Property & Epochs & Batch size & Hidden size & Hidden layers & Attention heads & Layers in head & Dropout & Train R2 & Validation R2 & Test R2 \\[6pt]
\hline
Radius of Gyration & 5 & 4 & 256 & 8 & 8 & 2 & 0.15 & 0.9605 & 0.9854 & 0.9529 \\
Heat Capacity & 5 & 4 & 256 & 8 & 8 & 2 & 0.15 & 0.9723 & 0.9607 & 0.9569 \\
Decorrelation Time & 5 & 4 & 256 & 8 & 8 & 2 & 0.15 & 0.9748 & 0.9543 & 0.9560 \\
\hline
Radius of Gyration & 5 & 4 & 512 & 8 & 8 & 2 & 0.15 & 0.9577 & 0.9754 & 0.9795 \\
Heat Capacity & 5 & 4 & 512 & 8 & 8 & 2 & 0.15 & 0.9709 & 0.9570 & 0.9631 \\
Decorrelation Time & 5 & 4 & 512 & 8 & 8 & 2 & 0.15 & 0.9680 & 0.9662 & 0.9607 \\
\hline
Radius of Gyration & 5 & 4 & 256 & 16 & 16 & 1 & 0.15 & 0.9608 & 0.9833 & 0.9523 \\
Heat Capacity & 5 & 4 & 256 & 16 & 16 & 1 & 0.15 & 0.9719 & 0.9724 & 0.9655 \\
Decorrelation Time & 5 & 4 & 256 & 16 & 16 & 1 & 0.15 & 0.9728 & 0.9774 & 0.9654 \\
\hline
Radius of Gyration & 5 & 4 & 256 & 16 & 16 & 2 & 0.15 & 0.9667 & 0.9879 & 0.9889 \\
Heat Capacity & 5 & 4 & 256 & 16 & 16 & 2 & 0.15 & 0.9794 & 0.9733 & 0.9689 \\
Decorrelation Time & 5 & 4 & 256 & 16 & 16 & 2 & 0.15 & 0.9806 & 0.9709 & 0.9687 \\
\hline
\end{tabular}
\end{adjustbox}
\caption{\label{tab:experiments}Different experiments conducted with various hyperparameters, each followed by $R^2$ values for the train, validation, and test sets.}
\end{table}

\section{Contribution of sequence length in predicting properties}

To assess the contribution of sequence length in predicting properties, as well as to examine the importance of sequence information, we conducted a modified training procedure. We modified the dataset so that for each protein sequence, we retained the length but replaced the actual sequence with a string of 'A's, effectively removing any meaningful sequence information. The performance of the model trained on this modified dataset is reported in Table \ref{tab:length_analysis}.

We observed that the R² scores were significantly lower, indicating that while length does have some correlation with the properties, it is insufficient to achieve high prediction R² scores. This highlights the importance of the specific sequence information in making accurate predictions.

\begin{table}[t!]
\begin{adjustbox}{width=0.65\columnwidth,center}
\begin{tabular}{|c|c|c|c|}
\hline
Property & Train R2 & Validation R2 & Test R2 \\
\hline
Radius of Gyration & 0.5184 & 0.7337 & 0.7436 \\
Heat Capacity & 0.5365 & 0.5636 & 0.6448 \\
Decorrelation Time & 0.2519 & 0.2188 & 0.3384 \\
\hline
\end{tabular}
\end{adjustbox}
\caption{\label{tab:length_analysis}Training, Validation, and Testing $R^2$ values for Radius of Gyration, Heat Capacity, and Decorrelation Time when the amino acid sequences were replaced with a string of 'A's.}
\end{table}

\section{Protein taxonomy of the dataset}

Similarity search of the sequences in the dataset was performed on the UniProt database, and the taxonomy information of each of the similarity hits were noted in a csv file. This taxonomy information was obtained using the NCBI Entrez API available within the BioPython library. For each of the similarity hits, the UniProt ID, the similarity fraction, and the taxonomy information is provided in the csv file named “taxonomy\_results.csv”.